\documentclass[a4paper]{jpconf}
\usepackage{graphicx}
\usepackage{comment}
\usepackage{amsmath}

\newcommand{\bT}{\bar{T}}

\newcommand{\kk}{\boldsymbol{k}}

\newcommand{\kb}{\mathbf{k}}

\newcommand{\xx}{\boldsymbol{x}}

\newcommand{\hc}{\text{h.c.}}

\newcommand{\TT}{\text{TT}}
\newcommand{\dd}{\text{d}}

\newcommand{\bra}[1]{\langle#1|}

\newcommand{\ket}[1]{|#1\rangle}

\newcommand{\nppp}{\nonumber\\[3pt]}

\begin{document}

\title{Quantum dynamics of bound states under spacetime fluctuations}

\author{Teodora Oniga and Charles H-T Wang}

\address{Department of Physics, University of Aberdeen, King's College, Aberdeen AB24~3UE, United Kingdom}

\ead{t.oniga@abdn.ac.uk, c.wang@abdn.ac.uk}

\begin{abstract}
With recent developments in high-precision quantum measurements, the question of whether observations of decoherence from spacetime fluctuations are accessible experimentally arises. Here we investigate the dynamics of bound states interacting with an environment of gravitons under the Markov approximation. The corresponding Lindblad master equation is presented that enables gravitational decoherence and dissipation due to zero-point spacetime fluctuations to be analyzed. Specifically, we consider a one-dimensional cavity of massless scalar particles that models a light beam with negligible spin polarizations being reflected between two free masses. Numerical simulations have been performed to illustrate the wave-modal dependent decoherence and dissipation of such a configuration. We further demonstrate the existence of nontrivial collective effects akin to superradiance, providing amplifications of gravitational decoherence for a large number of identical bosonic particles.
\end{abstract}

\section{Introduction}

Over the past few decades, quantum decoherence has become an ever more important field of physics. Given the universal nature of the matter-gravity interaction, spacetime fluctuations constitute an inescapable environment which may cause decoherence to occur in physical systems. Such effects may have an impact of high-precision measurements \cite{amelino, schiller, pikovski1, lamine, pfister} and astronomical observations \cite{amelino3} and, in addition, may also play a role in the classical appearance of macroscopic objects \cite{schlosshauer,giulini}. Thus it is important to understand the implications of environmentally induced gravitational decoherence. In the existing literature, this has been addressed mostly through phenomenological models, with different causes of decoherence, such as semiclassical metric fluctuations \cite{wang, breuer1, power}, quantized Newtonian gravity \cite{kay}, a thermal spacetime foam \cite{garay}, weak gravity \cite{anastopouloshu, blencowe} etc. These approaches have typically been limited by their specific requirements, including the Markov approximation, the use of single non-relativistic particles and assuming classical stochastic spacetime fluctuations.

Recently, a new theory of decoherence due to the entanglement of bosonic matter with spacetime fluctuations was developed from first principles \cite{oniga1, oniga2}. This approach is based on Dirac's constraint quantization of linearized gravity coupled with matter that allows to derive a gauge invariant influence functional technique used to obtain a general gravitational master equation for the reduced density matrix of the matter system by averaging over the environmental degrees of freedom. It was found that when this master equation is applied to a broad class of free bosons including scalar particles, for long times and with the Markov approximation, the nonunitary part of the master equation vanishes, suppressing completely decoherence. However, while the decoherence terms in the master equation disappear for free fields, for bound fields this may not be the case. Furthermore, it has been suggested that when a large number of identical particles is considered, nontrivial collective decoherence phenomena occur. Motivated by the above developments, in this paper we will apply the formalism in \cite{oniga1, oniga2} to scalar particles confined in a finite box, with discrete modes and zero boundary conditions, where we are indeed able to observe decoherence. The  scalar particles provide a useful initial description of the behaviours of other bosonic particles in a similar configuration, which is physically analogous to photons inside a reflective cavity when spin polarizations are negligible.

In the following we will use the relativistic units with the speed of light $c=1$ and rescale the gravitational constant $G \rightarrow Gc^4$ and the reduced Planck constant $\hbar \rightarrow \hbar/c$ as well as other physical quantities accordingly.  Time derivatives will be denoted by an overdot  $\dot {}$ and Hermitian conjugates by an superscript dagger ${}^\dag$. Spatial indices running from $1$ to $3$ are denoted by Latin letters and spatial and temporal indices running from $0$ to $3$ are denoted by Greek alphabets. Coordinates are given by $x^\mu=(x^0, x^1, x^2, x^3)= (x^0, \xx)= (x)$ where the zeroth coordinate is time. When indices are repeated, a summation is implied. The background metric is given by the Minkowski metric $\eta_{\mu \nu} = \text{diag}(-1, 1, 1, 1)$ .

\section{The gravitational master equation}

Let us consider an ensemble of environmental gravitons in a Gaussian state weakly interacting with a system of bosonic particles. We will further assume that the gravitational field follows a Planck distribution $N(\omega)= 1/(e^{{\hbar \omega}/{k_B T}}-1)$ at temperature $T$, with graviton frequency $\omega$ and the Boltzmann constant $k_B$, although in principle it is not required that the gravitational field state be thermal and as such, the same formalism can be used to investigate decoherence for a Gaussian state environment with a general distribution function $N(\omega)$.  The interaction picture evolution of the reduced density matrix of the bosonic system is then given by the exact master equation \cite{oniga1}:
\begin{eqnarray} \nonumber
\frac{d}{d t} \rho(t)
&=&  -\frac{i}{\hbar} [  H_{\text{self}}(x), \rho(t)] -
\frac{8 \pi G}{\hbar}
\int\! \frac{\dd^3 k}{2(2\pi)^3k} \\
&&
\Big \{
\big[
\tau^\dag_{ij} (\kk,t),\,
\tilde{\tau}_{ij}(\kk,t) \rho(t)
\big]
+
N(\omega_k)
\big[
\tau^\dag_{ij} (\kk,t),\,
\big[
\tilde{\tau}_{ij}(\kk,t),\,
\rho(t)
\big]\big]
+\hc
\Big\}.
\label{maseqn}
\end{eqnarray}
The gravitational self-interaction of the matter field is described by the Hamiltonian
\begin{eqnarray}
H_\text{self}
&=&
-2G\,\int\dd^3 x\,\dd^3 x'
\frac{T^{\mu\nu}(\vec{x},x^0)\bT_{\mu\nu}(\vec{x}',x'^0)}{|\vec{x}-\vec{x}'|}
\end{eqnarray}
where $T^{\mu\nu}$ is the stress energy tensor of the system of interest and $\bT_{\mu\nu}$ is the trace-reversed stress energy tensor. Here we have denoted by $\tau_{ij}(\kk,t)$ the normal ordered and particle number preserving stress energy tensor of the system of interest, in momentum space and in the transverse-traceless (TT) gauge. The TT projection operator
\begin{eqnarray}
P_{ijkl}
&=&
\frac12 P_{ik}P_{jl} + \frac12 P_{il}P_{jk} - \frac12 P_{ij}P_{kl}
\label{Pijkl}
\end{eqnarray}
where we have the transverse projection operator $P_{ij} = \delta_{ij}-\frac{\partial_i \partial_j}{\partial_k \partial_k}$ gives the TT stress tensor $\tau_{ij}(\kk,t)=P_{ijkl}(\kk) T^{kl}(\kk,t)$. The time integration from the initial time taken to be $0$ to a final time $t$ is denoted by:
\begin{eqnarray}
\tilde{\tau}_{ij}(\kk,t)
&=&
\int_{0}^{t} d t'
e^{-i \omega_k (t - t')}
\tau_{ij}(\kk,t')
\label{nt}.
\end{eqnarray}
As the gravitational self-interaction Hamiltonian part of the master equation corresponds to a unitary time evolution, it does not cause decoherence. Therefore, from now on we will only consider the dissipative part of the master equation.

\section{Discrete mode scalar field master equation}

A massive scalar field with dispersion relation $\omega_k^2 = k^2 + \mu^2$ and nonnegative $\omega_{k}$ has the following expression for its field operator in position space:
\begin{eqnarray}
\phi(\xx,t)
&=&
\int\dd^3k\,
\sqrt{\frac{\hbar}{2(2\pi)^3\omega_{k}}}\;
\big[a_{\kk}e^{ikx}+a_{\kk}^\dag e^{-ikx}\big].
\label{phiexp}
\end{eqnarray}
The spatial components of the stress energy tensor
\begin{eqnarray}
T_{ij}
&=&
\phi_{,i}\phi_{,j}
+
\frac12\delta_{ij}
(\phi_{,0}\phi_{,0}-\phi_{,k}\phi_{,k}-\mu^2\phi^2)
\end{eqnarray}
lead to the transverse-traceless stress energy tensor and its Hermitian conjugate
\begin{eqnarray}
\tau_{ij}(\kk,t)
&=&
P_{ijkl}(\kk)\int\dd^3k'
\frac{\hbar}{2\sqrt{\omega_{k'}}}
k'_k k'_l
\nppp&&
\left[
\frac{1}{\sqrt{\omega_{\kk'-\kk}}}
a^\dag_{\kk'-\kk} a_{\kk'}
e^{-i(\omega_{\kk'}-\omega_{\kk'-\kk})t}
+
\frac{1}{\sqrt{\omega_{\kk'+\kk}}}
a^\dag_{\kk'} a_{\kk'+\kk}
e^{-i(\omega_{\kk'+\kk}-\omega_{\kk'})t}
\right]
\label{stij1}
\\
\tau_{ij}^\dag(\kk,t)
&=&
P_{ijkl}(\kk)\int\dd^3k'
\frac{\hbar}{2\sqrt{\omega_{k'}}}
k'_k k'_l
\nppp&&
\left[
\frac{1}{\sqrt{\omega_{\kk'-\kk}}}
a^\dag_{\kk'} a_{\kk'-\kk}
e^{i(\omega_{\kk'}-\omega_{\kk'-\kk})t}
+
\frac{1}{\sqrt{\omega_{\kk'+\kk}}}
a^\dag_{\kk'+\kk}  a_{\kk'}
e^{i(\omega_{\kk'+\kk}-\omega_{\kk'})t}
\right].
\label{stij2}
\end{eqnarray}
The operators $a_{\kk}$ and $a_{\kk}^{\dag}$ are time-independent and satisfy the following commutation relations:
\begin{eqnarray}
\big[
a_{\kk}, a_{\kk'}^\dag
\big]
&=&
\delta(\kk-\kk')
\label{com1}
\\[3pt]
\big[
a_{\kk}, a_{\kk'}
\big]
&=&
\big[
a_{\kk}^\dag, a_{\kk'}^\dag
\big]
=
0.
\label{com3}
\end{eqnarray}
In a box of volume $V$ with zero boundary conditions, such that particles are reflected at the boundary, with $\tau_{ij}(\xx,t)=0$ if $\xx$ is outside the box, in terms of discrete momenta $\kb$:
\begin{eqnarray}
\tau_{ij}(\kb,t)
&=&
\int_V \dd^3x\, \tau_{ij}(\xx,t)  e^{-i\kb\cdot\xx}.
\end{eqnarray}
We can express the stress energy tensor in terms of continuous momenta $\kk$ as
\begin{eqnarray}
\tau_{ij}(\kk,t)
&=&
\int_V \dd^3x\, \tau_{ij}(\xx,t)  e^{-i\kk\cdot\xx}
\nppp
&=&
\sum_{\kb}
\Delta(\kk-\kb)\tau_{ij}(\kb,t)
\label{tkkb}
\end{eqnarray}
where
\begin{eqnarray}
\Delta(\kk)
=
\frac1V\int_V \dd^3 x \; e^{-i \kk \xx}
\label{Delta}
\end{eqnarray}
with the limit $ \lim_{V\to\infty}V\Delta(\kk)=(2\pi)^3\delta(\kk)$.

\section{Decoherence of an idealized light beam cavity}

As a one-dimensional reduction of the above discussion,
it is interesting to consider next in some detail
a situation analogous to a light beam bouncing between two end mirrors. Physically, these mirrors may be considered to be attached to classical free masses that average out metric fluctuations and so do not move in the TT coordinates currently adopted. The spin polarizations of the light are not taken into account by modelling the light as a massless scalar field.
We then have the following field operator inside the light beam with $0 < x < L$, $0<y <\Delta y$ and $0<z <\Delta z$:
\begin{eqnarray}
\phi(x,t)
&=&
\sum_{n}
\sqrt{\frac{\hbar}{L\omega_{n}}}\,
\big(
a_{n} e^{-i\omega_{n} t}
+
a_{n}^\dag  e^{i\omega_{n} t}
\big)
\sin\frac{\pi n x}{L}
\label{phiexpdis01}
\end{eqnarray}
where $V=L^3$ and $n_1,n_2,n_3=1,2,\cdots$ such that
\begin{eqnarray}
\kb
&=&
\left(\frac{\pi n_1}{L}, \frac{\pi n_2}{L},\frac{\pi n_3}{L}\right).
\label{kn}
\end{eqnarray}
This leads to the following creation and annihilation operators $a_{n}$ and $a_{n'}^\dag$:
\begin{eqnarray}
[a_{n},a_{n'}^\dag]
=
\delta_{n,n'},\quad
[a_{n},a_{n'}]
=0=
[a_{n}^\dag,a_{n'}^\dag].
\label{aacomm01}
\end{eqnarray}
Inside the light beam, the stress energy tensor takes the following effective 3-dimensional form:
\begin{eqnarray*}
T_{ij}^{\TT}(\xx, t)
&=&
\frac{1}{\Delta y\Delta z}P_{ij11} \phi_{,x}(\xx, t)\phi_{,x}(\xx, t).
\end{eqnarray*}
For a small beam width limit with $\Delta y, \Delta z \ll L \sim 1/k$ and by normal ordering and neglecting the terms of $T_{ij}^{\TT}(\kk, t)$ not conserving particle numbers, we obtain $\tau_{ij}(\kk, t)$:
\begin{eqnarray}
\tau_{ij}(\kk, t)
&=&
P_{ij11}(\kk)
\sum_{n,n'}
F(n,n',\kk)\,
a_{n'}^\dag a_{n} e^{-i(\omega_{n}-\omega_{n'}) t}
\label{t10a}
\\
\tau_{ij}^\dag(\kk, t)
&=&
P_{ij11}(\kk)
\sum_{n,n'}
F^*(n,n',\kk)\,
a_{n}^\dag a_{n'} e^{i(\omega_{n}-\omega_{n'}) t}
\label{t10b}
\end{eqnarray}
with
\begin{eqnarray}
F(n,n',\kk)
:=
\frac{2\pi^2 n\, n' \hbar}{L^3\sqrt{\omega_{n}\omega_{n'}}}\,
\int_0^L \dd x \,
\cos\frac{\pi n x}{L}
\cos\frac{\pi n' x}{L}\,
e^{-i k x \cos\theta}
\label{Func}
\end{eqnarray}
where the continuous wave vector $\kk$ of modulus $k$ forms a small angle $\theta$ with the $x$-axis. For the time evolution of this system neglecting memory effects, we can apply the Markov approximation. The time variable in the integration is changed from $t'$ to $s$ using $t'=t-s$, taking the limit $\int_{0}^{t} \dd s \to \int_{0}^{\infty} \dd s$. After applying the Sokhotski-Plemelj theorem $\int_0^\infty \dd s\, e^{-i\epsilon s} = \pi \delta(\epsilon) - i \,\mathrm{P} \frac{1}{\epsilon}$ where $\mathrm{P}$ denotes the Cauchy principal value, we can neglect the terms containing $\mathrm{P}$ which do not give rise to decoherence to obtain:
\begin{eqnarray}
\tilde{\tau}_{ij}(\kk,t)
&=&
\int_{0}^{\infty}\dd s\,
\tau_{ij}(\kk,t')
e^{-i k s}
\nppp
&=&
{\pi}P_{ij11}(\kk)\sum_{n \ge n'}
F(n,n',\kk)\,
a_{n'}^\dag a_{n}
e^{-i(\omega_{n}-\omega_{n'}) t}
\,\delta(k-\omega_{n}+\omega_{n'}).
\label{t10f}
\end{eqnarray}
After applying the customary rotating wave approximation and adopting the following notation:
\begin{eqnarray*}
\omega(n,\Delta n)
&=&
\omega_{n+\Delta n}-\omega_{n}
\\
\sum_{\underline{n,m,\Delta n,\Delta m}}
&=&
\sum_{\substack{{n,m,\Delta n,\Delta m}\\
{\omega(n,\Delta n)=\omega(m,\Delta m)}}}
\end{eqnarray*}
where $\kk(n,\Delta n)$ denotes $\kk$ with $k=\omega(n,\Delta n)$ and $\Delta n,\Delta m \ge 0$, we arrive at the master equation in Lindblad form \cite{oqs}:
\begin{eqnarray}
\frac{\dd}{\dd t} \rho(t)
&=&
\sum_{\underline{n,n',\Delta n,\Delta n'}}
\Gamma(n,n',\Delta n,\Delta n')
\nppp
&& \Big[
(1+N(\omega(n,\Delta n)))\big(
A(n',\Delta n')
\rho
A^\dag(n,\Delta n)
-
\frac12\{
A^\dag(n,\Delta n)
A(n',\Delta n'),
\rho
\}
\big)
\nppp&&
+
N(\omega(n,\Delta n)))\big(
A^\dag(n,\Delta n)
\rho
A(n',\Delta n')
-
\frac12\{
A(n',\Delta n')
A^\dag(n,\Delta n)),
\rho
\}
\big)
\Big].
\label{maseqtube2}
\end{eqnarray}
in terms of the operators $A(n,\Delta n)=a_{n}^\dag a_{n+\Delta n}$ and with the transition rate coefficient
\begin{eqnarray}
&&\Gamma(n,n',\Delta n,\Delta n')
=
\frac{G}{2\pi\hbar}
\omega(n,\Delta n)\, \nppp
&& \hspace{15pt}
\int\! \dd\Omega(\kk(n,\Delta n))
P_{11}^2(\kk(n,\Delta n))
F^*(n+\Delta n,n,\kk(n,\Delta n))
F(n'+\Delta n',n',\kk(n,\Delta n))
\label{Gtube0}
\end{eqnarray}
related directly to the rate of decoherence and dissipation.

To investigate the influence of zero-point spacetime fluctuations on a photon reflected by the boundaries of this light beam, let us consider a massless scalar master equation at zero gravitational temperature with $N(\omega)=0$. For massless particle $\Delta n=\Delta n'$ and the resulting transition rate coefficient $\Gamma(n,n',\Delta n)$ takes the form
\begin{eqnarray}
\Gamma(n,n',\Delta n)
&=&
\frac{4 \pi^3 \hbar\, G}{L^3}\,
\hat\Gamma(n,n',\Delta n)
\label{Gtube3}
\end{eqnarray}
in terms of the dimensionless coefficient
\begin{eqnarray}
\hat\Gamma(n,n',\Delta n)
&=&
\Delta n
{\sqrt{n(n+\Delta n)}}\,
{\sqrt{n'(n'+\Delta n)}}\,
\int_{-1}^{1}\! \dd\sigma\,(1-\sigma^2)^2
\nppp&&
\hspace{-55pt}
\int_0^1 \dd s \,
\cos(\pi n s)\,
\cos(\pi (n+\Delta n) s)\,
e^{i {\pi \sigma\Delta n} s }
\int_0^1 \dd s' \,
\cos(\pi n' s')\,
\cos(\pi (n'+\Delta n) s')\,
e^{-i {\pi \sigma\Delta n} s' }.
\nppp
\label{hGtube3}
\end{eqnarray}

As an illustrative numerical example in units where $c=1$ and ${4 \pi^3 \hbar\, G}/{L^3}=1$ displayed in the Fig. 1, we take a one-photon system with density matrix $\rho_{n,n'}(t)=\bra{n}\rho\ket{n'}(t)$ truncated for modes with $1 \le (n,n') \le 20$ and hence $1 \le \Delta n \le 19$. As shown, the initial density matrix at $t=0$ has a chosen 2-dimensional Gaussian profile centred at $(n,n')=(10,10)$. Under the nonunitary evolution, its centre moves towards the ground state with $(n,n')=(1,1)$ due to energy dissipation through spontaneous emissions of gravitons, with a gradually reduced off-diagonal width as a result of decoherence.

Fig. 2 shows a similar evolution where the photon density matrix is initially diagonal, having no superposition between different $n$-modes, with constant diagonal elements up to the truncation mode $n = 20$. The diagonality is dynamically preserved. It can clearly be seen that the higher decay rates for larger $n$-modes cause the formation of a peak off the smaller $n$-modes  while this peak settles towards the ground state at a decelerated pace.

\begin{figure}[ht!]
\begin{center}
\includegraphics[width=1\linewidth]{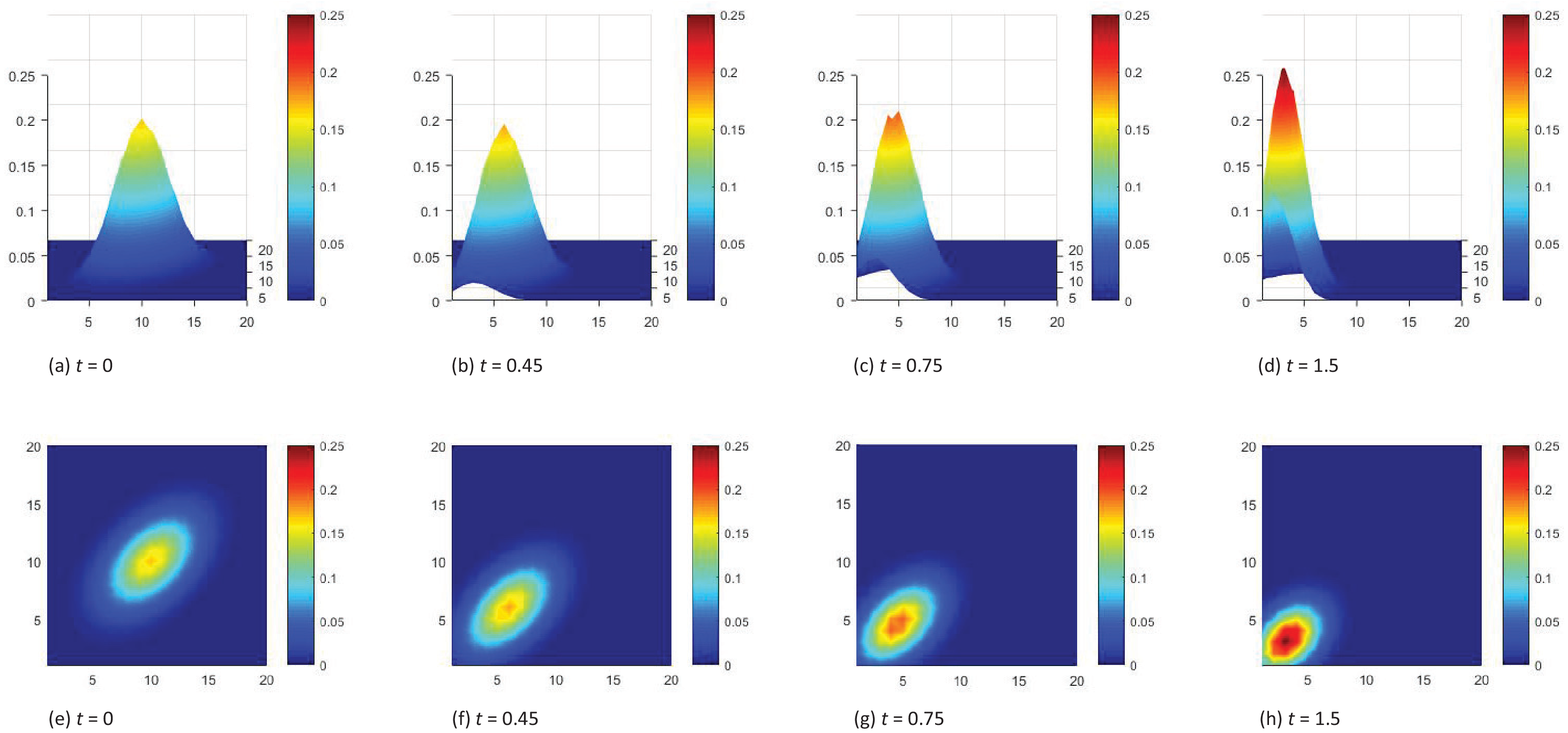}
\end{center}
\begin{small}
{\bf Figure 1.}
As explained in the main text, under the nonunitary evolution, the centre of the density matrix profile with an initial Gaussian shape moves towards the ground state due to energy dissipation, with a gradually reduced off-diagonal width as a result of decoherence. The evolution of the cavity photon beam density matrix $\rho_{n,n'}(t)$ is represented in side views (a)--(d) and in top views (e)--(h) where the matrix value is the height and matrix indices $(n,n')$ are along the horizontal axes.
\end{small}
\newline
\end{figure}

\begin{figure}[ht!]
\begin{center}
\includegraphics[width=1\linewidth]{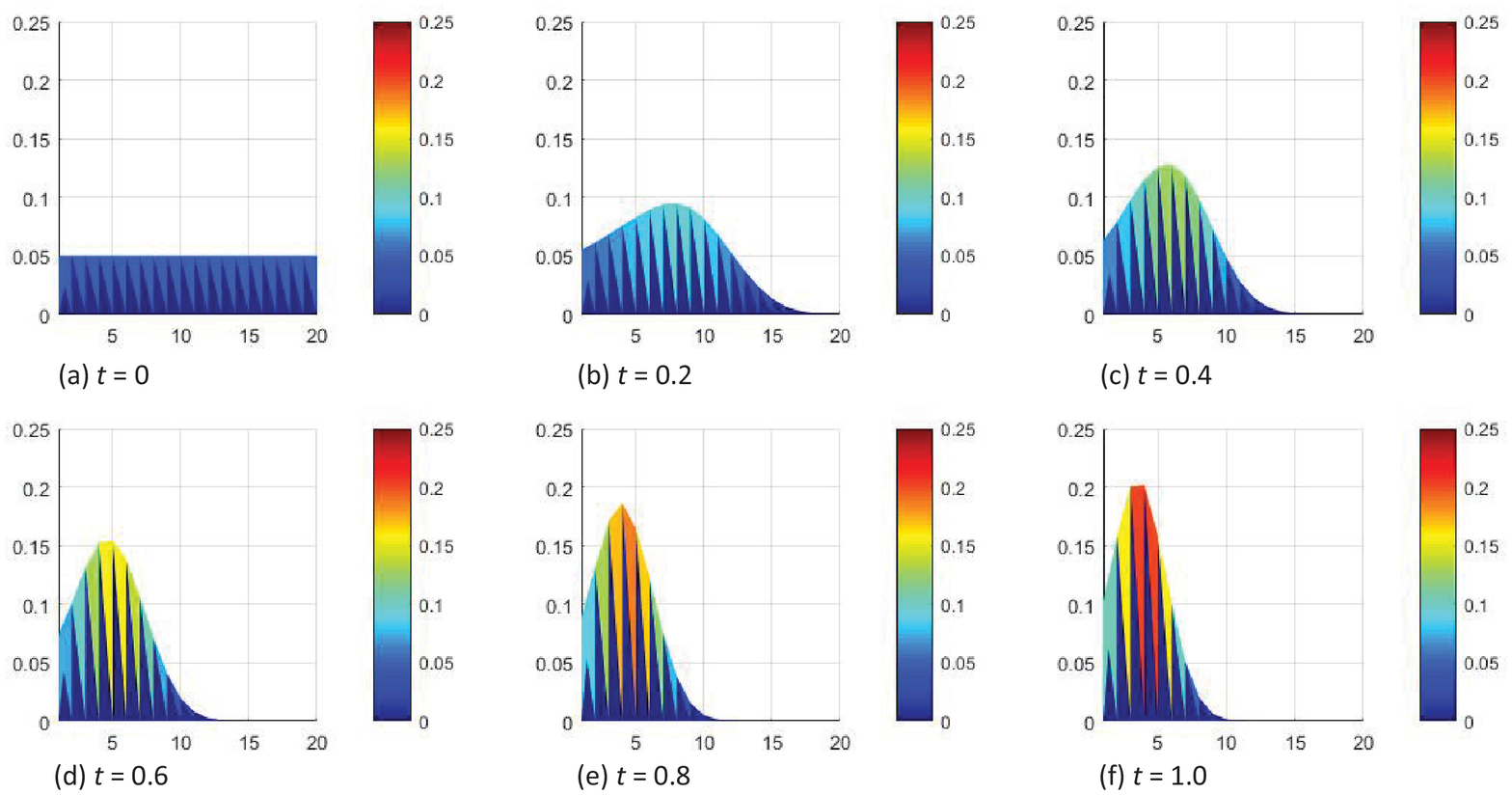}
\end{center}
\begin{small}
{\bf Figure 2.}
Here plots (a)--(f) demonstrate the evolution of the photon density matrix, which is initially diagonal, with constant diagonal elements $=0.05$ for $1 \le n \le 20$. It can clearly be seen that the higher decay rates for larger $n$-modes cause the formation of a peak (rather like traffic congestion) off the smaller $n$-modes while this peak settles towards the ground state in a slowing fashion.
\end{small}
\end{figure}

\section{Collective decoherence and radiance processes}

We would now like to consider the collective quantum dynamics of massless scalar particles at zero gravitational temperature using the master equation \eqref{maseqtube2} with $N(\omega)=0$ and transition rate coefficient $\Gamma(n,n',\Delta n)$ given by \eqref{Gtube3}. To start with, for one particle states given by $\ket{m}=a_m^\dag\ket{0}$, the density matrix will decay into lower energy states as $\Gamma(n,n, \Delta n)>0$. Let us look at a diagonal density matrix that at some point in time takes the form $
\rho
=\sum_m\rho(m)\ket{m}\bra{m}$
normalized so that $
\sum_m\rho(m)=1$. Then we have
\begin{eqnarray*}
\dot\rho
&=&
\sum_{m}\sum_{n<m}
\rho(m)
\Gamma(n,n,m-n)
\big[
\ket{n}\bra{n}-\ket{m}\bra{m}
\big]
\end{eqnarray*}
with the matrix elements given by
\begin{eqnarray*}
\bra{m}\dot\rho\ket{m}
&=&
\sum_{m'}\sum_{n<m'}
\rho(m')
\Gamma(n,n,m'-n)
\bra{m}\big[
\ket{n}\bra{n}-\ket{m'}\bra{m'}
\big]\ket{m}
\nppp
&=&
-
\sum_{m'}\sum_{n<m'}
\rho(m')
\Gamma(n,n,m'-n)
\delta_{m,m'}
+
\sum_{m'}\sum_{n<m'}
\rho(m')
\Gamma(n,n,m'-n)
\delta_{m,n}
\nppp
&=&
-
\sum_{n<m}
\rho(m)
\Gamma(n,n,m-n)
+
\sum_{n>m}
\rho(n)
\Gamma(m,m,n-m).
\end{eqnarray*}
Note that the density matrix stays diagonal throughout the time evolution. For comparison, let us also consider a many particle state, given by $N$ particles found in the $\ket{m}$, normalized such that $\ket{N_m} = \frac1{\sqrt{N_m!}}(a_m^\dag)^N\ket{0}$, with a multiple particle diagonal density matrix $\rho
=\ket{N_m}\bra{N_m}$ at a given time. Then, at that time, the master equation becomes:
\begin{eqnarray*}
\dot\rho
&=& 
\frac{N_m^2}{N_m!}
\sum_{n<m}
\Gamma(n,n,m-n)
a_{n}^\dag (a_m^\dag)^{{N_m}-1}
\ket{0}\bra{0}
(a_m)^{{N_m}-1}  a_{n}
\nppp&&
-
\frac{{N_m}}{N_m!}
\sum_{n<m}
\Gamma(n,n,m-n)
(a_m^\dag)^{N_m}
\ket{0}\bra{0}
(a_m)^{N_m}
\nppp&&
-
\frac{{N_m}({N_m}-1)}{2N_m!}
\sum_{n<m}
\Gamma(m,n,m-n)
a_{2m-n}^\dag
a_{n}^\dag (a_m^\dag)^{{N_m}-2}
\ket{0}\bra{0}
(a_m)^{N_m}
\nppp&&
-
\frac{{N_m}({N_m}-1)}{2N_m!}
\sum_{n<m}
\Gamma(n,m,m-n)
(a_m^\dag)^{N_m}\ket{0}\bra{0}
(a_m)^{{N_m}-2} a_{n}  a_{2m- n}.
\end{eqnarray*}
This reduces to the above one particle case when $\rho = \ket{m} \bra{m}$. However, we can see that for many particle states we have addition factors equal to the number of particles in the state, so that we have a collective, i.e. superradiant-like, decay of elements of the density matrix of the form:
\begin{eqnarray}
\bra{N_m}\dot\rho\ket{N_m}
&=&
-N_m
\sum_{n<m}
\Gamma(n,n,m-n).
\end{eqnarray}
More specifically, as with the description of electromagnetic superradiance \cite{dicke, gross}, the above relation shows the rate of spontaneous emission of gravitons by $N$ identical particles in the same state $\ket{m}$ increases by a factor of the particle number $N$ while the radiation intensity is increased by a further factor of $N$ resulting in an $N^2$ intensity amplification.

To investigate collective decoherence, we now consider a superposition state
\begin{eqnarray*}
\ket{\psi} = \frac1{\sqrt{2}}(\ket{N_m}+\ket{N_{m'}})
\end{eqnarray*}
with $m\neq m'$ and
\begin{eqnarray}
\rho
&=&
\ket{\psi}\bra{\psi}
\nppp
&=&
\frac1{2}\big(
\ket{N_{m}}\bra{N_{m}}
+\ket{N_{m}}\bra{N_{m'}}
+\ket{N_{m'}}\bra{N_{m}}
+\ket{N_{m'}}\bra{N_{m'}}
\big).
\label{rNN}
\end{eqnarray}
Therefore, we have arrived at a collective effect on the decoherence of the superposition of states $\ket{N_m}$ and $\ket{N_m}$ described by
\begin{eqnarray}
\bra{N_{m'}}\dot\rho\ket{N_m}
&=& 
-\frac14\big[
N_{m'}
\sum_{n<m'}
\Gamma(n,n,m'-n)
+
{N_{m}}
\sum_{n<m}
\Gamma(n,n,m-n)
\big]
\end{eqnarray}
where the decay rate of the off-diagonal element $\bra{N_{m'}}\rho\ket{N_m}$, increases with the particle numbers of the superposition states.

\section{Conclusion}
Motivated by recent theoretical and experimental developments in gravitational decoherence and quantum measurements, we have carried out an investigation of the open quantum dynamics of certain bound states subject to a fluctuating gravitational environment. Their non-free nature allows such systems to be formulated in terms of relatively well-understood Lindblad master equations, through which the gravitational decoherence and dissipation due to zero-point spacetime fluctuations have been analyzed. In a simplified resemblance to a (gravitationally noisy) laser interferometer, we then consider a one-dimensional cavity of massless scalar particles that models a light beam with negligible spin polarizations being reflected between two free masses. Numerical simulations have been performed to illustrate the wave-modal dependent decoherence and dissipation of such a configuration where a strong wave-modal dependence has been identified. Finally, we demonstrate the existence of novel collective quantum effects akin to superradiance originally formulated with electromagnetic interactions. This new collective process may eventually provide important amplifications of normally weak gravitational decoherence for a large number of identical bosonic particles.

\ack
The authors are grateful for hospitality to Martin Land and other Organizers of the IARD 2016 Conference, where a related lecture was delivered by TO. This work was supported by the Carnegie Trust for the Universities of Scotland (TO) and by the EPSRC GG-Top Project and the Cruickshank Trust (CW).

\end{document}